\documentclass[letterpaper,twocolumn,10pt]{article}
\usepackage{deperated/usenix}

\usepackage[english]{babel}
\usepackage{graphicx}
\usepackage{amsmath}
\usepackage{multirow}
\usepackage{tikz}
\newcommand*\circled[1]{\tikz[baseline=(char.base)]{%
    \node[shape=circle,draw,inner sep=0.6pt] (char) {\small #1};}}
\usepackage{xspace}
\usepackage{enumitem}
\usepackage{algpseudocode}
\usepackage{float}
\usepackage[font=small,labelfont=bf]{caption}
\usepackage[linesnumbered,ruled,noend]{algorithm2e}
\usepackage{booktabs}
\usepackage{makecell}
\usepackage[normalem]{ulem}
\usepackage{url}
\usepackage{xspace}

\setitemize{noitemsep,topsep=3pt,parsep=0em}

\DisableLigatures[f]{encoding=*, family=*}

\makeatother
\newcommand{\sysname}{LUMEN\xspace}
\newcommand{\para}[1]{\smallskip\noindent\textbf{#1}}

\DontPrintSemicolon
\SetAlgoCaptionSeparator{ }
\SetKwInput{KwInput}{Input}
\SetKwInput{KwOutput}{Output}

\makeatletter
\def\@maketitle{\newpage
 \vbox to 1.25in{
 \vskip 0in
 \begin{center}%
  {\Large\bf \@title \par}%
  \vskip 0.10in minus 0.05in
  {\large
   \lineskip .5em
   \begin{tabular}[t]{c}\@author
   \end{tabular}\par}%
 \end{center}%
 \par
 \vspace*{\fill}
 }
}
\makeatother

\begin{document}

\date{}
\title{\Large \bf \sysname: Coordinated Failure Recovery for Distributed LLM Serving}
\author{%
  Zhang Cao\textsuperscript{1}, Shujie Han\textsuperscript{2}, Juncheng Zhang\textsuperscript{1},
  Yuanming Ren\textsuperscript{1}, Yongkun Li\textsuperscript{3}, Patrick P. C. Lee\textsuperscript{1}\\[6pt]
  {\it \textsuperscript{1}The Chinese University of Hong Kong \quad
  \textsuperscript{2}Northwestern Polytechnical University}\\
  {\it \textsuperscript{3}University of Science and Technology of China}
}

\maketitle

\begin{abstract}
Modern large language model (LLM) serving clusters distribute inference requests across multiple worker processes on different GPUs, but failures are prevalent at scale. When a worker fails, the cluster simultaneously loses the failed worker's GPU-resident key-value (KV) caches and serving capacity, leaving surviving workers to absorb the redirected traffic while re-running interrupted requests from scratch. Existing fault-tolerant systems either restart interrupted requests from scratch or restore KV caches from checkpoints stored on a fixed neighboring worker, but both approaches route recovery work without considering current cluster load and leave the recovering worker idle during model reload. We present \sysname, a fault-tolerant LLM serving system that treats recovery as a load-aware coordination problem across three decision points: checkpoint placement before failures, interrupted-request distribution at failure time, and serving capacity restoration during model reload. We evaluate \sysname using both prototype experiments and large-scale simulations and demonstrate significant improvements in serving and recovery times.
\end{abstract}

\begin{sloppypar}
\section{Introduction}
\label{sec:intro}

Large language models (LLMs) are widely deployed as latency-sensitive online services, including conversational assistants, code completion, and real-time content generation, that must sustain interactive response times under continuous load \cite{kwon23, zheng24, yu22}. Since frontier models \cite{brown20, touvron23} exceed the memory of a single GPU, production deployments shard model copies using tensor or pipeline parallelism across large-scale GPU clusters spanning more than 10,000 GPUs to scale serving capacity \cite{jiang24, kokolis25, zu24}, with each model copy independently served by a process called {\em worker}. Unfortunately, failures are prevalent in large GPU clusters at deployment scale. Production traces report multiple hardware incidents per day, and LLM serving jobs encounter failures every few hours on average \cite{jiang24, kokolis25, zu24, hu24}. These failures originate at the GPU, node, or inter-node network layer, but they all manifest at the serving layer as one or more workers that become unavailable.

Existing failure recovery strategies in LLM serving fall into either {\em Stop-and-Restart} or {\em Fixed-Checkpointing}. Stop-and-Restart, the default in modern serving systems \cite{kserve,tgi,triton,vllmk8s}, reloads the failed worker and dispatches its interrupted requests to surviving workers, where each re-runs requests from scratch to rebuild the lost KV cache. This requires no checkpointed state, but imposes expensive replay costs on long-context requests and overwhelms already-stressed surviving workers. Our analysis (\S\ref{sec:motivation}) shows that in recovering a single-worker failure, Stop-and-Restart increases the mean time-to-first-token (TTFT) by 4.0$\times$ and the mean time-per-output-token (TPOT) by 1.6$\times$ across the surviving cluster, and the increases persist at scale from 4 to 64 workers. Fixed-Checkpointing, exemplified by D\'ej\`aVu \cite{strati24}, avoids full replays by streaming each request's KV cache to a predetermined worker as a checkpoint holder before any failure occurs; after a failure, all interrupted requests are restored from that worker. However, the checkpoint holder is chosen statically rather than by current load, so recovery work could be concentrated on a heavily loaded worker regardless of its queue depth. Both approaches also leave the recovering worker idle until the full model is reloaded. These limitations motivate {\em load-aware} recovery that uses current load status across workers to guide three recovery decisions: (i) where to place KV checkpoints before failures, (ii) where to route interrupted requests at failure time, and (iii) how to utilize the recovering worker during model reload.

We present \sysname, a fault-tolerant, load-aware recovery system for LLM serving. Our key insight is that fast recovery from worker-level failures can be formulated as a load-aware coordination problem. \sysname comprises three coordinated mechanisms: (i) {\em load-aware KV checkpointing}, which places each request's KV checkpoint on the worker expected to have the least recovery load, spreading checkpoints across the cluster rather than concentrating them on one fixed neighbor; (ii) {\em locality-aware recovery scheduling}, which routes each interrupted request to its checkpoint holder, redirecting only requests with small checkpointed prefixes when a holder would still be overloaded; and (iii) {\em speculation-assisted progressive recovery}, which extends speculative decoding \cite{leviathan23, chen23} for recovery by loading a lightweight draft model on the recovering worker while the full model is reloaded in the background, thereby enabling the recovering worker to immediately contribute temporary capacity through speculative decoding without staying idle.

We implement and evaluate \sysname as a prototype built on SGLang \cite{zheng24} and as a large-scale simulator built on Vidur \cite{agrawal24}. On a four-worker prototype serving Qwen3-32B \cite{yang25qwen3}, \sysname reduces mean TTFT by 44.4\% and 7.1\%, mean TPOT by 15.9\% and 7.0\%, and recovery time by 50.0\% and 34.9\% over Stop-and-Restart and Fixed-Checkpointing, respectively; on an eight-worker prototype serving Qwen3-14B \cite{yang25qwen3}, \sysname reduces mean TTFT by 29.6\% and 15.9\%, mean TPOT by 7.1\% and 4.2\%, and recovery time by 64.1\% and 63.9\% over the same baselines. Large-scale simulations confirm \sysname's performance gains for cluster sizes of up to 64 workers. We will open-source \sysname's prototype and simulator in the final paper.

\section{Background}
\label{sec:background}

\subsection{LLM Serving Basics}

\noindent{\bf Prefill, decode, and KV cache.} LLM inference follows an autoregressive Transformer model \cite{vaswani17}. A request first runs a {\em prefill}, which processes all prompt tokens in parallel and generates the first output token. To bound per-step compute and interleave prefill with ongoing decodes, modern LLM serving systems split long prompts into fixed-size chunks (e.g., 1,024 tokens per chunk) and process them across iterations \cite{agrawal24sarathi}. The model then enters the {\em decode} phase, which generates one token per step, conditioned on the prompt and all previously generated tokens \cite{zhong24, patel24}. To avoid recomputing attention over the full prefix at every decode step, the system caches per-token keys and values as the {\em KV cache}. Following paged KV management \cite{kwon23}, we refer to a fixed-size block of cached keys and values for contiguous token positions as a {\em KV page}. The KV cache grows with sequence length and consumes a large fraction of GPU memory \cite{liu24cache, qin25}. Resuming a request after a failure without re-running prefill requires preserving these KV pages.

\para{Speculative decoding.} As the decode phase is memory-bandwidth-bound, GPU compute units often sit idle while waiting for model weights and KV cache. {\em Speculative decoding} \cite{leviathan23, chen23} turns linear decoding into parallel verification: a small {\em draft model} rapidly proposes a sequence of draft tokens, and the large {\em target model} verifies them in a single forward pass, effectively performing a mini-prefill step. Token acceptance is sequential: all draft tokens up to the first rejected position are accepted, and the target model contributes one additional corrected token at the rejection point; remaining draft tokens are discarded. As the target model verifies multiple tokens in parallel, it increases throughput without changing the output distribution \cite{leviathan23}. Efficiency depends on the draft model's acceptance rate and its per-step overhead relative to the target-model decode step.

\para{Parallelism.} Production deployments of large models shard one model copy across multiple GPUs using two complementary schemes: {\em tensor parallelism (TP)} shards each layer across GPUs within a node and issues fine-grained collectives every decode step; {\em pipeline parallelism (PP)} partitions layers into stages placed on different GPUs or nodes and exchanges only activation tensors between stages, at the cost of pipeline bubbles and more complex cross-stage state management. A {\em worker} is a complete copy of the model weights that can independently serve requests; in practice, a worker may occupy one GPU, a TP group within one node, a PP group across stages, or a combined TP+PP group. The GPUs inside a worker are tightly coupled: the failure of any one GPU disables the entire worker.

\subsection{Failures in Distributed LLM Serving}

\noindent{\bf Failure model.}
Production-serving deployments share the same GPU hardware and interconnect stack as training clusters and are hence exposed to the same hardware and network failures. These failures originate at three physical layers. (i) {\em GPU-level failures} come from ECC errors, overheating shutdown, PCIe/NVLink errors, driver resets, CUDA runtime aborts, or out-of-memory kills; losing any worker process halts the entire TP or PP group. (ii) {\em Node-level failures} come from a host crash, kernel panic, power event, or planned reboot; they disable all GPUs and serving processes on the node simultaneously, taking down all co-located workers at once. (iii) {\em Inter-node network failures} come from NIC failures, link flaps, switch port errors, fabric partitions, or persistent stragglers; they interrupt cross-worker activation exchanges, forcing the affected worker to be declared unavailable once collective latency violates the service-level objective (SLO).

All three failure types cause one or more workers to become unavailable, which we call a {\em worker-level failure}. The unavailable worker is the {\em failed worker}, which loses its GPU-resident KV cache, while the cluster loses that worker's prefill and decode capacity until model reload completes. In this work, we model each failure conservatively as requiring a full model reload, which is the dominant cost for node-level and persistent GPU-level failures, and address worker-level failures regardless of their underlying cause.

\para{Recovery flow.} When a worker-level failure occurs, the workers that remain available are the {\em surviving workers}; the failed worker after restart, while reloading its model, is the {\em recovering worker}. We focus on two recovery baselines (\S\ref{sec:intro}): {\em Stop-and-Restart}, which re-runs prefill from scratch for every interrupted request on surviving workers, and {\em Fixed-Checkpointing}, which restores interrupted requests from KV checkpoints stored on a fixed neighboring worker, which we call the request's {\em checkpoint holder}. Both baselines leave the recovering worker idle until its model finishes reloading.

\section{Motivation}
\label{sec:motivation}

We characterize the recovery cost of Stop-and-Restart under worker failures to
identify the root causes that \sysname targets. We select Stop-and-Restart as
the primary baseline as it represents the default deployment
\cite{kserve,tgi,triton,vllmk8s}. Also, it isolates the costs of lost KV cache
and serving capacity without introducing the load imbalance associated with
Fixed-Checkpointing.

\subsection{Impact of Worker Failures}
\label{subsec:failure_impact}

\begin{figure}[!t]
    \centering
    \begin{tabular}{@{\ }c@{\ }c@{\ }}
        \multicolumn{2}{c}{\includegraphics[width=0.95\linewidth]{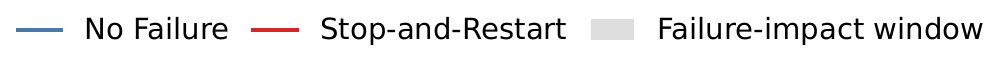}}
        \vspace{-6pt}\\
        \includegraphics[width=0.47\linewidth]{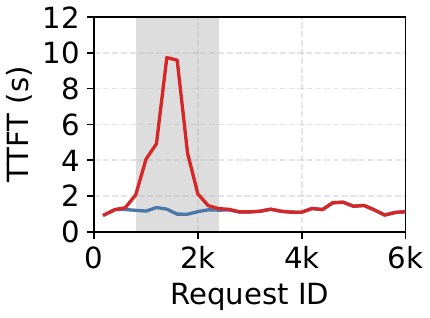} &
        \includegraphics[width=0.47\linewidth]{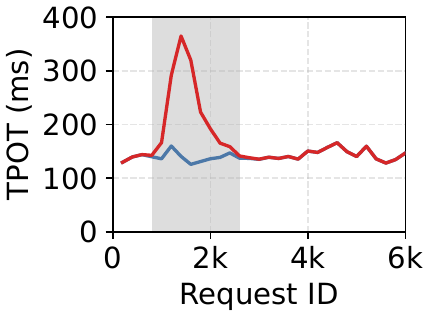} \\
        {\small \makecell[c]{(a) Mean TTFT}} &
        {\small \makecell[c]{(b) Mean TPOT}}
    \end{tabular}
    \vspace{-9pt}
    \caption{Single-worker failure in a simulated four-worker cluster.}
    \label{fig:failure_impact}
    \vspace{-6pt}
\end{figure}

\noindent{\bf Observation 1: A single-worker failure degrades both TTFT and TPOT cluster-wide.} We use our simulator to run Llama-3-70B \cite{llama3} on a cluster of four workers driven by the Splitwise-Conv trace \cite{patel24}. We evaluate Stop-and-Restart by failing a single worker after the cluster reaches steady state and compare it with the No-Failure case. Figure~\ref{fig:failure_impact} shows mean TTFT and TPOT in buckets across request IDs (with 200~requests per bucket), averaged over five runs, and marks the {\em failure-impact window} (i.e., the period from when the failure occurs until the cluster recovers to normal performance); see the failure-impact window definition and detailed cluster setup in \S\ref{subsec:eval_methodology}. Within the failure-impact window, the mean TTFT increases from 1.16\,s to 4.69\,s (4.0$\times$) and the mean TPOT increases from 138.9\,ms to 224.6\,ms (1.6$\times$). Per-request traces show that only 2.7\% of requests in this window are in flight on the failed worker; the remaining 97.3\% run on surviving workers. Thus, the failure degrades latency cluster-wide, affecting not only the failed worker's own requests but also requests on surviving workers. Once the recovering worker rejoins, both metrics return to their No-Failure levels.

\begin{figure}[!t]
    \centering
    \begin{tabular}{@{\ }c@{\ }c@{\ }}
        \multicolumn{2}{c}{\includegraphics[width=0.75\linewidth]{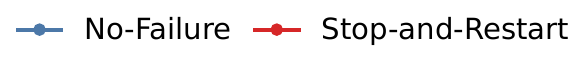}}
        \vspace{-6pt}\\
        \includegraphics[width=0.47\linewidth]{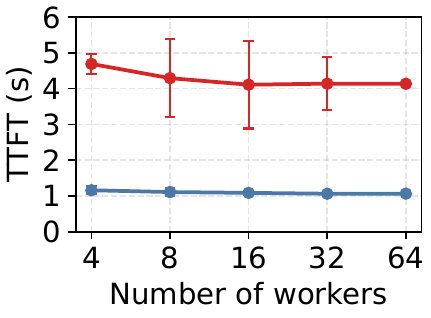} &
        \includegraphics[width=0.47\linewidth]{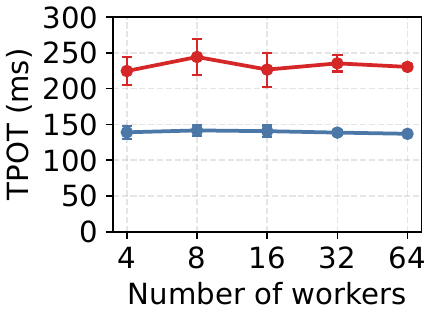} \\
        {\small \makecell[c]{(a) Mean TTFT}} &
        {\small \makecell[c]{(b) Mean TPOT}}
    \end{tabular}
    \vspace{-9pt}
    \caption{Worker failures versus cluster size.}
    \label{fig:mot_wscale}
    \vspace{-6pt}
\end{figure}

\para{Observation 2: Failure-induced degradation persists at larger cluster sizes.}
We scale the simulated cluster from 4 to 64 workers, maintaining a fixed load of
1.4 QPS per worker to evaluate scalability.
After the cluster reaches
steady state, we fail 25\% of the workers (1 to 16 failed workers).
Figure~\ref{fig:mot_wscale} shows the mean TTFT and mean TPOT over the
failure-impact window for No-Failure and Stop-and-Restart. Stop-and-Restart
degrades latency by a similar factor at every cluster size: the mean TTFT stays
in 4.1--4.7\,s, about 4$\times$ the No-Failure case of around 1\,s, and the mean
TPOT stays near 230\,ms, about 1.6$\times$ No-Failure of around 140\,ms.

\subsection{Root Cause Analysis}
\label{subsec:root_cause}

We analyze the root causes of Observations~1 and~2 by categorizing requests in the failure-impact window. For {\em interrupted requests} (i.e., requests running on a failed worker), Stop-and-Restart re-prefills them from scratch using each request's retained token history to rebuild the lost KV cache. For {\em uninterrupted requests} (i.e., remaining requests, including those already running on surviving workers and newly arrived requests during model reload), they do not lose KV cache on failure but may queue behind redirected traffic. Within each uninterrupted request's TTFT, we measure the queueing delay before prefill starts (the residual time is the prefill execution time). Table~\ref{tab:source_breakdown} shows the breakdown.

\begin{table}[!t]
\centering
\caption{Breakdown of Stop-and-Restart. Values are mean $\pm$ 95\% confidence intervals over five runs.}
\label{tab:source_breakdown}
\vspace{-9pt}
\small
\begin{tabular}{cccc}
\toprule
\multirow{2}{*}{Workers} & \multicolumn{2}{c}{Uninterrupted TTFT (s)} & \multirow{2}{*}{\makecell[c]{Interrupted\\Replay TTFT (s)}} \\
\cmidrule(lr){2-3}
 & Total & Queueing & \\
\midrule
4  & 4.1\,$\pm$\,0.2 & 3.3\,$\pm$\,0.2 & 25.6\,$\pm$\,7.1 \\
8  & 3.8\,$\pm$\,1.1 & 3.0\,$\pm$\,1.0 & 24.0\,$\pm$\,1.4 \\
16 & 3.5\,$\pm$\,1.2 & 2.8\,$\pm$\,1.2 & 27.1\,$\pm$\,1.7 \\
32 & 3.5\,$\pm$\,0.7 & 2.7\,$\pm$\,0.6 & 29.4\,$\pm$\,2.9 \\
64 & 3.6\,$\pm$\,0.1 & 2.8\,$\pm$\,0.1 & 27.8\,$\pm$\,0.7 \\
\bottomrule
\end{tabular}
\vspace{-6pt}
\end{table}

\para{Observation 3: Model reloading delays uninterrupted requests due to capacity loss.}
While a failed worker reloads its model, its incoming traffic is redirected to
surviving workers. At a fixed 25\% failure rate, this increases the load on
every surviving worker by one-third regardless of cluster size (the lost 25\% of
load is absorbed by the remaining 75\% of workers, i.e., a one-third increase).
For TTFT, uninterrupted requests do not lose KV state but still wait behind
redirected traffic. Table~\ref{tab:source_breakdown} shows that their mean TTFT
stays within 3.5--4.1\,s at every cluster size, dominated by queueing delay
(78--80\%). For TPOT, the same capacity shortfall increases decode contention on
surviving workers, matching the roughly 1.6$\times$ TPOT degradation in
Figure~\ref{fig:mot_wscale}(b).

\para{Observation 4: Lost KV cache forces expensive replay for interrupted requests.} We measure {\em replay TTFT} for each interrupted request as the time from its original arrival to the first output token of the re-prefill attempt; it counts (i) the work already spent before the failure, (ii) any queueing after dispatch, and (iii) the full re-prefill required to rebuild the KV cache. The replay TTFT is within 24.0--29.4\,s across cluster sizes, roughly 5.9--8.4$\times$ the uninterrupted requests' TTFT. The per-request re-prefill cost is similar across cluster sizes, with only slight variation due to queueing behind redirected traffic on overloaded surviving workers.
Although interrupted requests are only 2.7\% of the failure-impact window (Observation~1), their high replay TTFT increases the overall window mean to 4.69\,s.

\para{Recovery objective.} We aim to minimize recovery costs cluster-wide. This
requires load-aware recovery decisions that (i) preserve KV checkpoints to
eliminate replay TTFT for interrupted requests, and (ii) recover capacity during
model reload to mitigate queueing delay (TTFT) and decode contention (TPOT) for
uninterrupted requests.

\subsection{Challenges}
\label{subsec:challenges}

Our recovery optimization faces three design challenges.

\para{Challenge~1: Checkpoint placement.}
Storing in-flight KV caches on other workers bypasses expensive re-prefills. However, Fixed-Checkpointing (\S\ref{sec:intro}) \cite{strati24} statically maps all requests from a failed worker to a single neighbor (i.e., the checkpoint holder). This creates severe recovery bottlenecks where queueing negates KV-reuse benefits. Alternatively, remote storage offloading \cite{qin25} avoids host memory pressure but suffers from unpredictable network latency during recovery. The challenge is dynamically distributing checkpoints to prevent recovery hotspots, while bounding host memory usage and keeping restore costs lower than recomputation in practice.

\para{Challenge~2: Recovery dispatch.} At failure time, routing interrupted
requests presents a trade-off between KV reuse and load balancing. Maximizing KV
reuse by sending all requests to their checkpoint holders creates severe
recovery hotspots. Conversely, rebalancing requests to underloaded workers
discards checkpointed KV pages, forcing expensive full re-prefills. The
challenge is to dynamically strike a balance by redirecting only requests with
minimal checkpointed prefixes away from overloaded holders, ensuring load
balance while bounding recomputation penalties.

\para{Challenge~3: Reload-window assistance.}
Recovering workers traditionally remain idle during lengthy model reloads, while surviving workers are bottlenecked by the decode throughput of redirected traffic \cite{patel24}. To extract usable decode throughput immediately, we
repurpose speculative decoding \cite{leviathan23, mcdanel25, liu25} for fault
tolerance: the recovering worker rapidly loads a lightweight draft model to
produce tokens for an overloaded surviving worker to verify. While this advances
multiple tokens per step to swiftly clear backlogs, it introduces two critical
difficulties. First, verifying stale drafts on an already-stressed worker
degrades performance; this overhead must therefore be bounded to drop stale
tokens without stalling the normal decode path. Second, the recovering worker
must seamlessly transition from draft assistance back to full service without
incurring another disk-loading stall at the end of the reload window.

\section{\sysname Design}
\label{sec:design}

\sysname addresses the challenges in \S\ref{subsec:challenges} via three coordinated mechanisms: {\em load-aware KV checkpointing}, {\em locality-aware recovery scheduling}, and {\em speculation-assisted progressive recovery}. All three mechanisms share a common principle that recovery decisions follow current load observations and direct extra work toward less-loaded workers.

\subsection{Design Overview}

\begin{figure}[!t]
\centering
\includegraphics[width=0.99\linewidth]{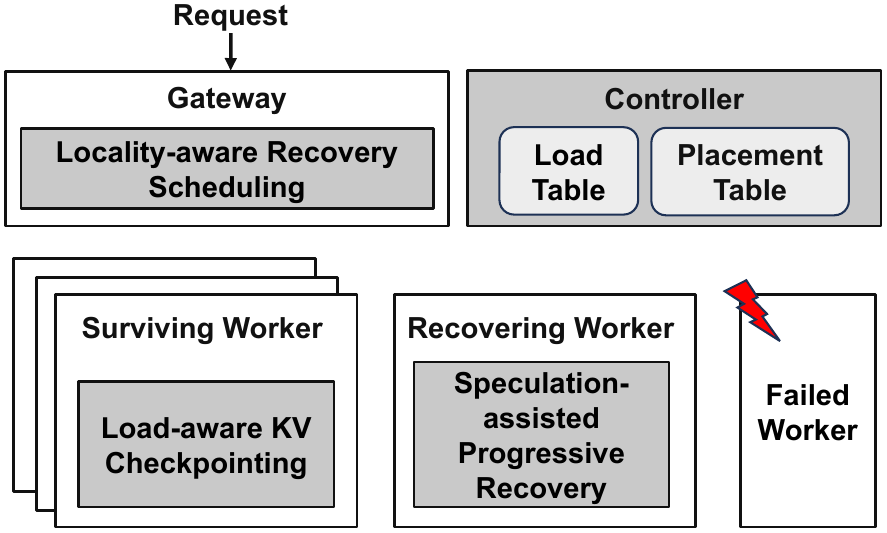}
\vspace{-6pt}
\caption{\sysname overview.}
\label{fig:arch}
\vspace{-6pt}
\end{figure}

\noindent{\bf Architecture.}
Figure~\ref{fig:arch} shows the \sysname architecture. In existing LLM serving systems (e.g., SGLang \cite{zheng24}), a {\em gateway} merely routes requests and triggers naive Stop-and-Restart recovery upon failure, while workers independently manage local GPU memory. \sysname augments the gateway to retain the {\em token history} of every in-flight request, including the prompt and accumulated output token IDs. It also introduces a centralized {\em controller}, which maintains a {\em load table} of per-worker queueing and capacity signals and a {\em placement table} that maps each request to its checkpoint holder; both tables drive checkpoint placement in steady state and request dispatch when a failure occurs.

\para{Workflow.} In steady state, load-aware KV checkpointing (\S\ref{subsec:checkpointing}) continuously copies each request's new KV pages to its checkpoint holder's CPU memory off the GPU critical path; the controller reads the load table to choose that checkpoint holder and records the choice in the placement table. When a worker fails, the controller triggers two parallel actions: locality-aware recovery scheduling (\S\ref{subsec:scheduling}), which dispatches each interrupted request to its checkpoint holder for KV reuse and redirects some requests away from overloaded checkpoint holders to avoid hotspots, and speculation-assisted progressive recovery (\S\ref{subsec:progressive}), which keeps the recovering worker useful while it reloads the target model.

\para{Goals.}
\sysname targets three design goals.
\begin{itemize}[leftmargin=*]
\item {\bf Fast recovery}, which lowers post-failure latency by jointly preserving in-flight KV cache and recovering lost serving capacity before target-model reload completes.
\item {\bf Low steady-state overhead}, which keeps the always-on checkpoint path off the GPU decode path and bounded in host memory so that failure-free TTFT and TPOT match those of a serving system without \sysname.
\item {\bf Scalability}, which preserves the recovery benefit across cluster sizes and across concurrent multi-worker failures.
\end{itemize}

\subsection{Load-aware KV Checkpointing}
\label{subsec:checkpointing}

To address Challenge~1, \sysname introduces load-aware KV checkpointing. Rather than routing all checkpoints to a static worker \cite{strati24}, it evaluates real-time load signals to distribute checkpoints across multiple workers at request granularity.

\begin{figure}[!t]
\centering
\includegraphics[width=0.9\linewidth]{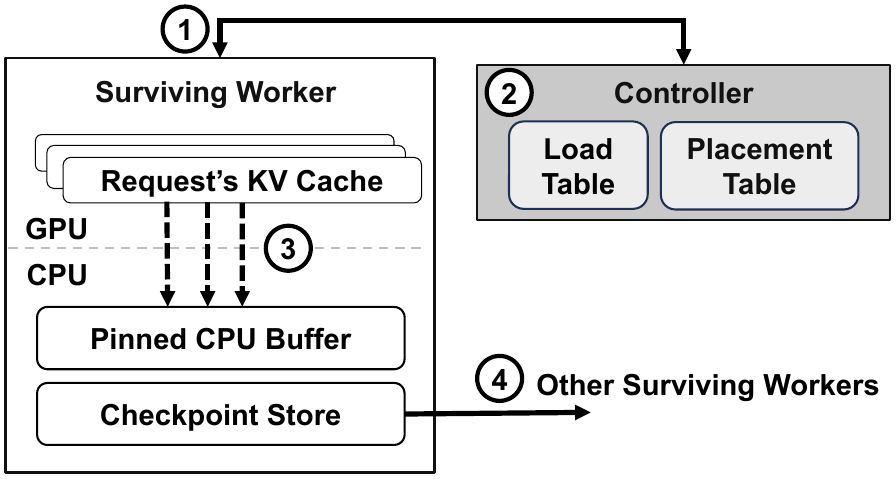}
\vspace{-6pt}
\caption{Load-aware KV checkpointing.}
\label{fig:kv_checkpointing}
\vspace{-6pt}
\end{figure}

\para{Checkpoint placement.}
Figure~\ref{fig:kv_checkpointing} shows \sysname's KV checkpoint placement. The controller aggregates per-worker feedback (\circled{1}) into a load table, which then drives the placement table (\circled{2}) to map each active request to its designated checkpoint holder. To estimate potential restore pressure, the load table tracks three per-worker states: available checkpoint-store capacity, queueing delay, and reserved KV checkpoint footprints. Rather than relying on periodic polling, \sysname employs event-driven updates: it adjusts reserved KV checkpoint footprints instantly upon allocation or release, and updates queue delays as workers complete requests.

Load-table updates impose minimal overhead by decoupling the controller from heavy data transfers. The controller operates at request granularity rather than token granularity and exchanges only lightweight metadata (e.g., checkpoint-holder IDs, request IDs, capacities, and queueing delays) during prefill and request completion. Consequently, control traffic scales with request throughput, independent of token generation or sequence length. Meanwhile, large KV pages stream directly peer-to-peer between workers, completely bypassing the controller.

\sysname designates a single worker to store the KV checkpoint for each active request. It requires that the designated worker's available host memory can accommodate the request's {\em reserved KV checkpoint footprint}, a conservative allocation derived from the maximum context length rather than the current dynamic KV size. By reserving this footprint upon assignment and releasing it upon completion, \sysname prevents checkpoint holders from exhausting host memory. Also, maintaining exactly one checkpoint per request bounds steady-state memory usage and simplifies placement logic. \sysname bounds checkpoint memory usage with a per-worker host-memory limit, using the same budget as Fixed-Checkpointing (e.g., D\'ej\`aVu \cite{strati24}); in our prototype deployments (\S\ref{subsec:eval_methodology}), this limit is 80\,GB per worker for Qwen3-14B and 160\,GB per worker for Qwen3-32B \cite{yang25qwen3}. Fallback mechanisms for checkpoint-holder failures are detailed in~\S\ref{subsec:discussion}.

\sysname scores each candidate worker by two metrics: (i) {\em queueing delay} $q_w$ measures the average wait time on worker $w$ from request arrival to prefill initiation; and (ii) {\em restore pressure} $p_w(r)$ estimates the expected mean KV restore latency for worker $w$ if request $r$ is assigned to it. To compute $p_w(r)$, the controller first averages the reserved KV footprints of the checkpoints that $w$ would hold after assigning $r$, then divides this average by a fixed host-to-GPU bandwidth \cite{gao24}. These metrics capture distinct latency dimensions: $q_w$ reflects current congestion, while $p_w(r)$ infers the future recovery load
concentrated on $w$ during a failure. \sysname balances them using a tunable weight $\lambda$ to accommodate variations in request rate, model size,
and hardware bandwidth.

\sysname assigns the KV checkpoint to the candidate worker that minimizes the weighted sum of these metrics. The candidate set $F(r)$ comprises all workers with sufficient storage to accommodate $r$'s reserved KV footprint, excluding the worker currently serving $r$. This physical separation ensures that no single-worker failure can compromise both a request's active KV cache and its checkpoint.
The resulting checkpoint holder, $h(r)$, is:
\begin{equation}
\label{eq:placement}
h(r) = \arg\min\nolimits_{w \in F(r)} \big( q_w + \lambda p_w(r) \big).
\end{equation}
We set $\lambda=1$ by default and evaluate its sensitivity (\S\ref{sec:eval}).
\sysname finalizes the checkpoint holder $h(r)$ upon prefill completion, logging this assignment in the placement table for the controller to route interrupted requests during recovery. Subsequently, all KV pages are continuously streamed to this designated checkpoint holder until the request terminates and its checkpoint is released. Notably, this placement strategy is intentionally asymmetric: it proactively shifts future recovery load toward workers with greater spare capacity by allocating them more KV checkpoints.

\para{Incremental checkpoint pipeline.} Figure~\ref{fig:kv_checkpointing} also shows that \sysname asynchronously offloads KV cache to checkpoint holders without blocking the GPU critical path. Each worker provisions two host-memory regions: a {\em pinned CPU buffer} for staging local GPU-to-network transfers, and a {\em checkpoint store} for persisting incoming remote KV pages for checkpointing. To minimize overhead, workers track checkpoint progress and transmit only newly generated KV pages after each prefill or decode batch. A background thread copies these pages from GPU memory to the local pinned buffer (\circled{3}), then streams them over the network to the designated checkpoint store (\circled{4}). This two-stage pipeline completely overlaps with normal inference execution.

To facilitate recovery, each transferred page carries a {\em page tag}: a hash value of the token IDs stored in page $i$, paired with the page's end position in the request token sequence (i.e., the cumulative token length after page $i$). Since this tag derives purely from the request's token sequence, any worker can independently reconstruct it using the token history retained by the gateway. To guarantee atomicity, pages become visible in the checkpoint store only upon complete reception; if a failure interrupts a transfer, the lookup simply halts at the last valid tag, leaving only the suffix after the last complete page for recomputation. Consequently, these page tags enable a recovering worker to pinpoint the longest contiguous checkpointed prefix (\S\ref{subsec:scheduling}).

\subsection{Locality-aware Recovery Scheduling}
\label{subsec:scheduling}

\sysname addresses Challenge~2 via locality-aware recovery scheduling. Upon failure detection by the gateway, the controller initially routes interrupted requests to their checkpoint holders to maximize KV reuse. To prevent these holders from becoming recovery hotspots, it dynamically sheds load by redirecting a subset of requests to less-loaded workers, deliberately trading KV reuse for recomputation.

\para{Recovery dispatch.} To prioritize KV reuse, the controller routes interrupted requests to their designated checkpoint holders. If a checkpoint holder has also failed, \sysname marks the request for recomputation and assigns it to the least-loaded surviving worker. While this locality-first approach preserves cache state, relying on historical load observations risks concentrating recovery traffic on specific workers, necessitating the subsequent rebalancing step.

\para{Load rebalancing.} To mitigate residual imbalance during recovery,
\sysname identifies overloaded checkpoint holders using an {\em average-based request-count rule}: it compares each surviving worker's total request load (i.e., queued, running, and newly assigned interrupted requests) against the cluster-wide average computed after the initial recovery dispatch. When a holder is overloaded, \sysname migrates its assigned requests to the least-loaded worker, prioritizing requests in increasing order of their actual checkpointed size (i.e., the volume of KV cache successfully persisted before the failure). This greedy heuristic minimizes recomputation cost: since migrating a request forfeits its local checkpoint and forces a full re-prefill, shedding requests with the smallest saved prefixes first limits the total lost state. The controller iteratively recomputes worker loads after each migration and always targets the most congested worker first, until no worker exceeds the cluster-wide average.

\para{KV-reuse recovery.} Following dispatch, checkpoint holders execute
in-place recovery by exploiting preserved KV caches. From the request's token
history, the holder deterministically regenerates page tags to locate the
longest contiguous checkpointed prefix within its local store, subsequently
loading the matching KV pages into GPU memory. If the persisted prefix is
incomplete relative to the current token history, \sysname bypasses full
recomputation by performing a partial prefill on the uncheckpointed
suffix. The restored KV pages serve as the context for this suffix prefill,
enabling the request to seamlessly resume autoregressive decoding.

\subsection{Speculation-assisted Progressive Recovery}
\label{subsec:progressive}

\begin{figure}[!t]
\centering
\includegraphics[width=0.9\linewidth]{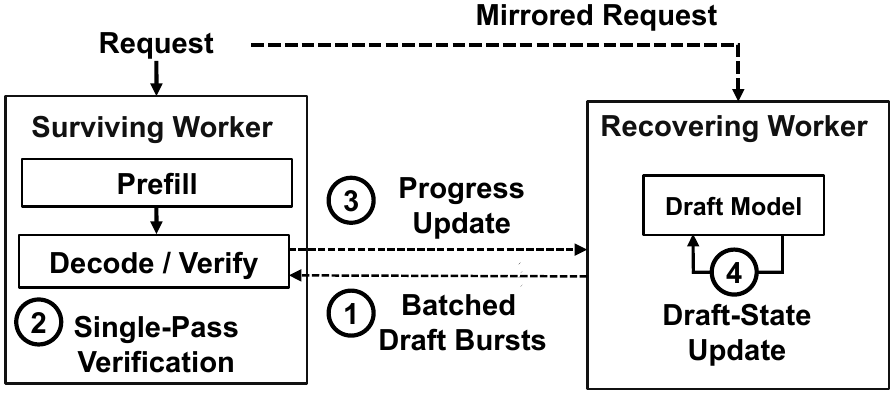}
\vspace{-6pt}
\caption{Speculation-assisted progressive recovery.}
\label{fig:progressive_workflow}
\vspace{-6pt}
\end{figure}

\sysname addresses Challenge~3 via speculation-assisted progressive recovery. The recovering worker moves through four internal states (Figure~\ref{fig:progressive_workflow}): \texttt{LOADING\_DRAFT} loads the draft model; \texttt{ASSIST} feeds draft tokens to one paired surviving worker while the target model loads in the background; \texttt{HOTSWAP} transfers the preloaded target-model weights to the GPU; and \texttt{FULL\_SERVICE} resumes normal target-model serving. This ensures draft assistance and target-model loading overlap, and the worker returns to full service without another disk-loading stall.

\para{Draft-worker pairing.} Upon entering the \texttt{ASSIST} state, \sysname pairs the recovering worker exclusively with the surviving worker exhibiting the highest average queueing delay (\S\ref{subsec:checkpointing}), ensuring draft assistance is directed to the most congested worker. One-to-one pairing consolidates all draft tokens into a single batch, generating sufficient volume to amortize the target model's verification overhead; unpaired surviving workers continue to receive new requests from the gateway transparently. To initiate assistance, the paired surviving worker transmits a {\em mirror request}, an exact token copy that seeds the draft model rather than producing user-facing output, for each in-flight request in its batch. The recovering worker continuously evaluates these mirror requests, accumulating unverified draft tokens until a target speculative depth of $K$ is reached, at which point the tokens are transmitted back as a single burst (\circled{1} in Figure~\ref{fig:progressive_workflow}). To minimize per-token coordination overhead, bursts completed within the same iteration are aggregated into a single network transfer.

\para{Single-pass verification.} A naive implementation requires two target-model forward passes per decode step: one to verify drafted requests and another to decode unassisted ones. \sysname circumvents this overhead by constructing a single fused batch per step (\circled{2} in Figure~\ref{fig:progressive_workflow}). To maintain structural consistency, \sysname enforces a uniform shape of $K+1$ sequence positions for every request: the first position holds the latest committed token, and the remaining $K$ positions contain either draft tokens (for assisted requests) or dummy {\em placeholders} (for unassisted ones). This uniformity enables the entire batch to execute as a single forward pass within a single CUDA graph. For unassisted requests, only the first position produces the next token, while placeholder outputs are discarded, mimicking a standard decode step. For assisted requests, the worker applies standard speculative decoding \cite{leviathan23}, accepting the longest valid prefix and appending the corrected token at the first rejection point. As one-to-one pairing concentrates draft tokens into this single batch, the vast majority of requests are assisted, rendering the computational overhead of placeholders negligible.

\para{Draft-state alignment.} To maintain asynchronous execution, the paired surviving worker never blocks the recovering worker. Instead, after each decode step, it transmits a progress update (\circled{3} in Figure~\ref{fig:progressive_workflow}) that includes the newly committed tokens and the output lengths for all mirror requests. Upon receiving this update, the recovering worker aligns its draft KV cache by sequence position (\circled{4} in Figure~\ref{fig:progressive_workflow}). Specifically, it compares the authoritative committed tokens against its local draft tokens; the draft KV remains valid only up to the first mismatched position. From that divergence point onward, the recovering worker truncates the invalid draft KV, replays the remaining committed tokens through the draft model to rebuild the context, and resumes speculative generation. Strict positional alignment is essential, as value-based matching is ambiguous due to token recurrence across different sequence positions.

\para{Background loading and final switch.} To circumvent prolonged synchronous stalls, \sysname initiates a background process to load target-model weights into host memory concurrently with the \texttt{ASSIST} state. This masks the high-latency disk-to-host I/O and isolates only the rapid host-to-GPU memory transfer onto the \texttt{HOTSWAP} critical path. Upon completion of host-side loading, the recovering worker discards its transient draft state, deallocates the draft runtime, and transfers the preloaded target weights to the GPU to resume \texttt{FULL\_SERVICE}. This transition is seamless: the authoritative request state and token history remain anchored at the surviving worker and gateway, respectively, so no state needs to be migrated at switch time. The pipeline also degrades gracefully under adverse conditions: cache misses fall back to re-prefill, lagging draft bursts are ignored by the surviving worker without stalling decode, and unexpected background loading delays simply extend the \texttt{ASSIST} phase until the hot-swap can proceed safely.

\subsection{Discussion}
\label{subsec:discussion}

\para{Trade-offs.} \sysname incurs negligible steady-state overhead, primarily consuming host memory and asynchronous network bandwidth for KV checkpointing (\S\ref{sec:eval}). Locality-aware scheduling may induce recomputation penalties by redirecting requests away from overloaded checkpoint holders; however, this cost is minimized by migrating requests with the shortest checkpointed prefixes. Speculation-assisted recovery introduces a transient draft-model footprint and bounded verification overhead, constrained by the one-to-one worker pairing. Finally, \sysname currently maintains a single KV checkpoint per request, falling back to full recomputation if the checkpoint holder itself fails; we defer multi-checkpoint replication to future work.

\para{Multi-worker failures.} \sysname consolidates concurrent multi-worker failures into a unified recovery event. The gateway aggregates all interrupted requests and executes locality-aware recovery scheduling across the surviving cluster. Concurrently, each recovering worker independently initiates speculation-assisted recovery, prioritizing pairings with surviving workers exhibiting the highest queueing delays. To prevent verification bottlenecks, the controller enforces strict one-to-one assignments: if multiple recovering workers target the same node, it reroutes the latter to the next most congested, unpaired worker. If all surviving workers are actively paired, any remaining recovering workers bypass draft assistance and directly load the full target model.

\para{Gateway and controller failures.} \sysname assumes the gateway and controller remain operational during recovery. Although logically centralized, both components can achieve high availability and fault tolerance via standard replication.

\section{Implementation}
\label{sec:impl}

\para{Prototype implementation.} We implement the \sysname prototype atop SGLang (v0.5.6)~\cite{zheng24} in Python ($\sim$13\,K LoC), leveraging PyTorch (v2.9)~\cite{pytorch}, FlashInfer (v0.5)~\cite{flashinfer}, ZeroMQ~\cite{zeromq}, and gRPC (v1.75)~\cite{grpc}. We integrate four core components into SGLang: (i) a centralized controller for checkpoint holder assignment; (ii) a ZMQ-based transport layer for streaming newly generated KV pages to remote host memory; (iii) a recovery-aware Sarathi-Serve scheduler~\cite{agrawal24sarathi} that partitions traffic into KV-reuse, recomputation, and new-request queues; and (iv) a progressive recovery pipeline that executes the draft model, synchronizes tokens via gRPC, and hot-swaps the target model upon background loading completion. Additionally, we augment the SGLang gateway with load-aware routing logic, parsing health checks to track worker states and isolating non-\texttt{FULL\_SERVICE} workers from new traffic, while explicitly permitting mirror requests for workers in the \texttt{ASSIST} state.

\para{Simulator implementation.} For large-scale evaluation, we extend Vidur~\cite{agrawal24} ($\sim$5\,K LoC) with five modules: (i) an event-driven worker state machine governing recovery transitions (\S\ref{subsec:progressive}); (ii) a global scheduler for post-failure request dispatch; (iii) a recovery-aware Sarathi-Serve scheduler~\cite{agrawal24sarathi} mirroring the prototype's queue abstraction; (iv) a KV-restore analytical model that replaces Vidur's default prefill-recomputation penalty with bandwidth-derived KV-load times; and (v) a KV-distribution module to execute load-aware checkpoint placement decisions across the cluster.

\begin{figure*}[!t]
\centering
\begin{tabular}{@{\ }c@{\ }c@{\ }c@{\ }c@{\ }}
\multicolumn{4}{c}{\includegraphics[width=0.75\linewidth]{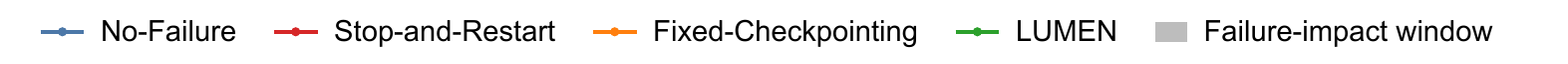}}
\vspace{-6pt}\\
\includegraphics[width=0.243\linewidth]{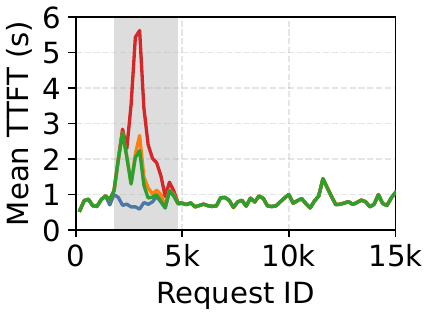} &
\includegraphics[width=0.243\linewidth]{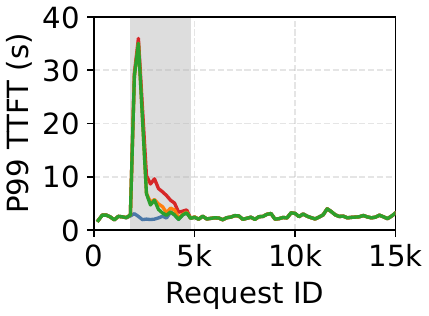} &
\includegraphics[width=0.243\linewidth]{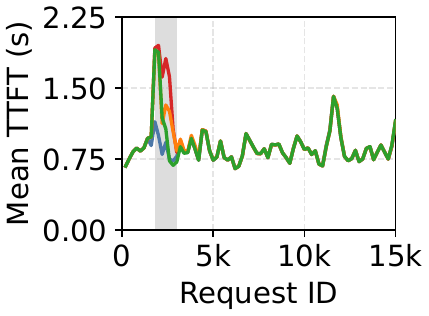} &
\includegraphics[width=0.243\linewidth]{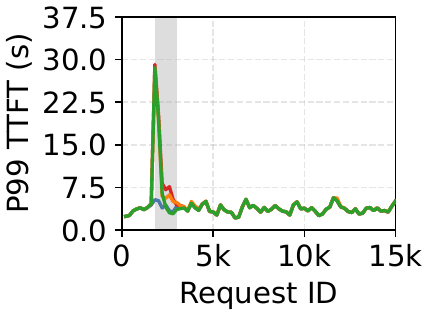} \\
{\small \makecell[c]{(a) Mean TTFT, 4-worker}} &
{\small \makecell[c]{(b) P99 TTFT, 4-worker}} &
{\small \makecell[c]{(c) Mean TTFT, 8-worker}} &
{\small \makecell[c]{(d) P99 TTFT, 8-worker}}
\vspace{3pt}\\
\includegraphics[width=0.243\linewidth]{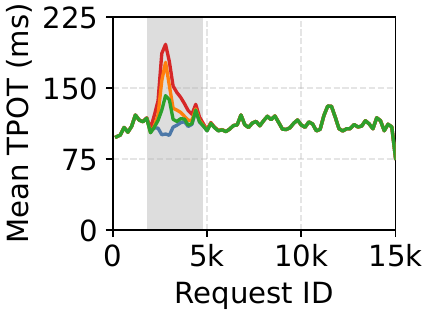} &
\includegraphics[width=0.243\linewidth]{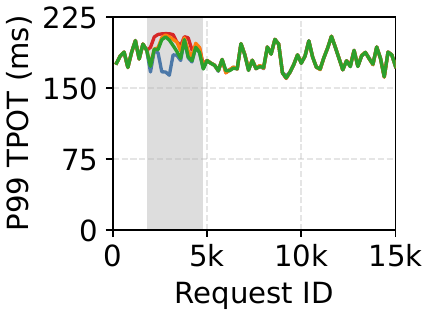} &
\includegraphics[width=0.243\linewidth]{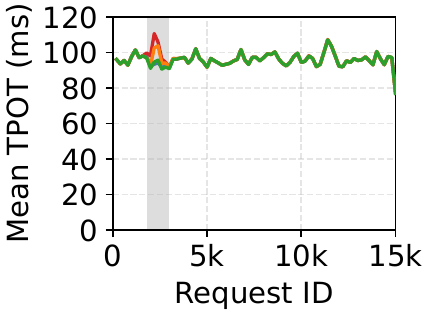} &
\includegraphics[width=0.243\linewidth]{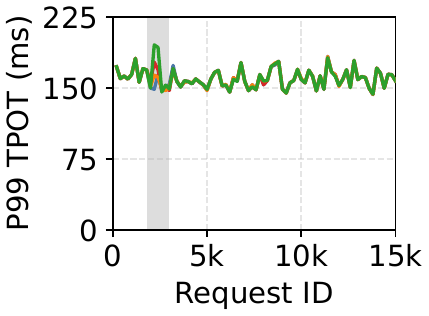} \\
{\small \makecell[c]{(e) Mean TPOT, 4-worker}} &
{\small \makecell[c]{(f) P99 TPOT, 4-worker}} &
{\small \makecell[c]{(g) Mean TPOT, 8-worker}} &
{\small \makecell[c]{(h) P99 TPOT, 8-worker}}
\end{tabular}
\vspace{-9pt}
\caption{Experiment~A.1: (Prototype) End-to-end recovery under a single-worker failure for four-worker (Qwen3-32B) and eight-worker (Qwen3-14B) deployments. We show the failure-impact window (in grey box) of Stop-and-Restart. Note that outside the failure-impact window (i.e., no failures), each request runs the same query workflow, so the curves overlap.}
\label{fig:proto_e2e}
\vspace{-6pt}
\end{figure*}

\section{Evaluation}
\label{sec:eval}

We evaluate \sysname using both a hardware prototype and a large-scale simulator: the prototype measures real system behavior (e.g., disk I/O, network transfer, and GPU memory management under failures), while the simulator enables controlled configurations across cluster sizes, failure counts, and load regimes beyond the reach of our physical testbed.

\subsection{Methodology}
\label{subsec:eval_methodology}

\noindent{\bf Evaluation environments.} The prototype runs on a cluster of four nodes. Each node has two Intel Xeon Platinum 8358P processors (32~cores each at 2.6\,GHz; 10~cores per socket allocated to our experiments), 200\,GB DRAM, two 894\,GB Samsung PM893 SATA SSDs, and two NVIDIA A800-SXM4-80 GB GPUs connected by NVLink at 200 GB/s inter-GPU bandwidth per GPU and attached to the host over PCIe~4.0~$\times$16; the nodes are interconnected by 10\,Gbps Ethernet. The cluster provides eight A800 GPUs in total. We evaluate two deployments: (i) four workers (each spanning the two GPUs of one node under tensor parallelism) serving Qwen3-32B with a Qwen3-4B draft model \cite{yang25qwen3}, and (ii) eight workers (one GPU per worker, two workers per node) serving Qwen3-14B with a Qwen3-1.7B draft model \cite{yang25qwen3}.

The simulator runs on a server with two Intel Xeon Gold 6330 processors (28~cores each at 2.0\,GHz) and 503\,GB DRAM. By default, the simulator models a deployment with ten workers (each spanning four A100-80 GB-class GPUs) serving Llama-3-70B with a Llama-3-8B draft model \cite{llama3} at a measured acceptance rate of 0.60. It uses 26\,GB/s GPU-to-CPU bandwidth \cite{gao24} and 2\,GB/s local-SSD bandwidth \cite{sheng23}.

\para{Workloads.} The prototype uses the ShareGPT trace \cite{sharegpt}, and the simulator uses the Splitwise-Conv trace \cite{patel24}; the two datasets exercise complementary operating regimes (longer, conversational sessions in ShareGPT versus shorter, high-concurrency prefill-decode phases in Splitwise-Conv).

\para{Default settings.} Unless otherwise stated, both environments use a speculative depth of $K=4$, the Sarathi-Serve scheduler \cite{agrawal24sarathi} with continuous batching \cite{yu22}, a chunk size of 1,024 tokens, and a batch cap of 512. Each setup issues 15,000 requests (40,000 for Experiments~B.4 and~B.5 over a longer trace) under Poisson arrival. The default request rates are 12\,QPS (four-worker), 10\,QPS (eight-worker), and 14\,QPS (simulator). We inject a single worker failure per deployment.

\para{Baselines.} We compare \sysname against three schemes: No-Failure (running the same workload without failure injection), Stop-and-Restart (S\&R), and Fixed-Checkpointing (F-Ckpt)
(both defined in \S\ref{sec:intro}).

\para{Metrics.} We bucketize a sequence of request IDs (200~requests per bucket) and measure the mean/P99 TTFT and TPOT per bucket. We identify the {\em failure-impact window} for each recovery scheme, starting when the bucketized mean TTFT exceeds that of the aligned No-Failure bucket by more than 5\%, and continuing until three consecutive buckets fall back within 5\%. We define the recovery time as the duration of the failure-impact window. We present average results over five runs, with 95\% confidence intervals under the Student's t-distribution shown in tables and bar charts; confidence intervals are omitted from line plots for brevity.

\subsection{Prototype Evaluation}
\label{subsec:eval_prototype}

\noindent{\bf Experiment~A.1 (End-to-end recovery).} Figure~\ref{fig:proto_e2e} shows bucketized mean and P99 TTFT and TPOT for all schemes under a single-worker failure. All three recovery schemes increase latency above No-Failure after the failure, but \sysname recovers fastest: it has the lowest peak latency and returns to the baseline first. Stop-and-Restart has the highest peak (mean TTFT above 5\,s at four workers, dropping below 1\,s before the failure), as it stops serving and recomputes the lost prefill from scratch, whereas Fixed-Checkpointing and \sysname keep the peak much lower by reusing the checkpointed KV cache. The failure increases latency more at four workers than at eight workers because it removes a larger share of serving capacity (25\% vs. 12.5\%).
The P99 metrics show the same trend with more noise; the comparisons below use failure-impact window means.

\begin{table}[!t]
\centering
\small
\caption{Experiment~A.1 (Prototype): Recovery time, with latency breakdown for interrupted (Int) and uninterrupted (Unint) requests under the eight-worker Qwen3-14B deployment.}
\label{tab:proto_e2e}
\vspace{-9pt}
\setlength{\tabcolsep}{2.5pt}
\begin{tabular}{@{}lc cc cc@{}}
\toprule
 & \multirow{2}{*}{\makecell[c]{Recovery\\(s)}} & \multicolumn{2}{c}{Mean TTFT (s)} & \multicolumn{2}{c}{Mean TPOT (ms)} \\
\cmidrule(lr){3-4}\cmidrule(lr){5-6}
Scheme & & Int & Unint & Int & Unint \\
\midrule
No-Failure        & --            & --                       & 0.83$\pm$0.23           & --                     & 92.6$\pm$2.1           \\
S\&R  & 83.3          & 12.5$\pm$3.6             & 1.3$\pm$0.48            & 116.5$\pm$6.6          & 99.3$\pm$3.1           \\
F-Ckpt & 82.8          & 10.7$\pm$3.1            & 1.1$\pm$0.37            & 59.7$\pm$9.6           & 97.2$\pm$2.6           \\
\sysname          & 29.9          & 11.0$\pm$3.4             & 0.87$\pm$0.26           & 58.0$\pm$12.4          & 93.1$\pm$2.3           \\
\bottomrule
\end{tabular}
\vspace{-6pt}
\end{table}

Table~\ref{tab:proto_e2e} decomposes the failure-impact window by interrupted and uninterrupted requests; we focus on the eight-worker deployment, while the results for the four-worker case are consistent. Interrupted requests are few (1.7\%); for example, their mean TTFT under Stop-and-Restart is 9.6$\times$ that of uninterrupted requests within the failure-impact window because the failed prefill is discarded and recomputed from scratch. As TTFT is measured from each request's original arrival time, the mean TTFT for interrupted requests varies significantly and is similar under \sysname and Fixed-Checkpointing; \sysname's advantage for this population is in TPOT rather than TTFT. \sysname reduces their mean TPOT by 50.2\% over Stop-and-Restart: resuming decode on restored KV cache already brings Fixed-Checkpointing to 59.7\,ms, and the speculation path reduces it further to 58.0\,ms. For the dominant uninterrupted requests (98.3\%), \sysname reduces mean TTFT by 33.1\% over Stop-and-Restart, confirming that fast recovery benefits a much larger set of requests that never run on the failed worker.

\sysname adds no measurable steady-state overhead before the failure: its curves overlap No-Failure on both TTFT and TPOT, with mean TTFT of 0.76\,s on the four-worker deployment and 0.79\,s on the eight-worker deployment (identical to No-Failure in both cases), and mean TPOT matching within 0.1\,ms. Since this end-to-end measurement already includes the controller's checkpoint placement and load-report handling, it confirms that the controller adds no measurable overhead at prototype scale.

Over the failure-impact window, \sysname achieves the largest latency reduction. It reduces mean TTFT by 44.4\% on the four-worker deployment and by 29.6\% on the eight-worker deployment, outperforming Fixed-Checkpointing in both cases. Mean TPOT drops by 15.9\% (four workers) and 7.1\% (eight workers) over Stop-and-Restart, because the speculation path handles decode while the recovering worker's target model is still loading. Furthermore, \sysname reduces recovery time by 50.0\% over Stop-and-Restart and by 34.9\% over Fixed-Checkpointing on the four-worker deployment, and by 64.1\% over Stop-and-Restart and by 63.9\% over Fixed-Checkpointing on the eight-worker deployment. Fixed-Checkpointing recovers only slightly faster than Stop-and-Restart despite reusing its checkpointed KV cache, as it sends all of a failed worker's restores to a single checkpoint holder that then absorbs both its own traffic and the entire restore load; \sysname avoids this concentration by spreading restore load across all surviving workers.

\begin{figure}[!t]
\centering
\begin{tabular}{@{\ }c@{\ }c@{\ }}
\multicolumn{2}{c}{\includegraphics[width=0.95\linewidth]{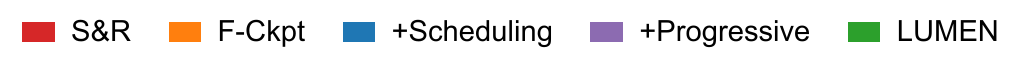}}
\vspace{-6pt}\\
\includegraphics[width=0.48\linewidth]{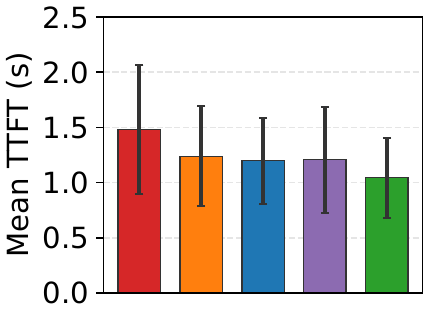} &
\includegraphics[width=0.48\linewidth]{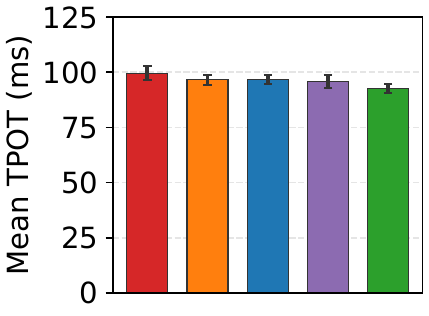} \\
{\small \makecell[c]{(a) Mean TTFT}} &
{\small \makecell[c]{(b) Mean TPOT}}
\end{tabular}
\vspace{-9pt}
\caption{Experiment~A.2 (Prototype): Recovery-path breakdown.}
\label{fig:proto_ablation}
\vspace{-6pt}
\end{figure}

\para{Experiment~A.2 (Recovery-path breakdown).} We decompose \sysname's recovery path to examine the contributions of individual techniques: {\bf +Scheduling} enables the KV-reuse path, including load-aware KV checkpointing (\S\ref{subsec:checkpointing}) and locality-aware recovery scheduling (\S\ref{subsec:scheduling}), so interrupted requests resume from their KV checkpoints; {\bf +Progressive} enables only the speculative path (\S\ref{subsec:progressive}) without KV reuse; and {\bf \sysname} is the full system. Figure~\ref{fig:proto_ablation} compares various schemes over the failure-impact window on the eight-worker deployment. +Scheduling alone reduces mean TTFT by 19.1\% by avoiding lost-prefix replay, whereas +Progressive alone reduces it by 18.5\%; both fall below Stop-and-Restart. For mean TPOT, the two approaches contribute complementarily: +Scheduling reduces mean TPOT to 96.7\,ms by resuming decode from restored KV pages, and +Progressive reduces it to 95.8\,ms by handling decode through the speculation path. \sysname combines both and achieves the lowest mean TTFT (1.0\,s) and mean TPOT (92.5\,ms).

\begin{figure}[!t]
\centering
\begin{tabular}{@{\ }c@{\ }c@{\ }}
\multicolumn{2}{c}{\includegraphics[width=0.95\linewidth]{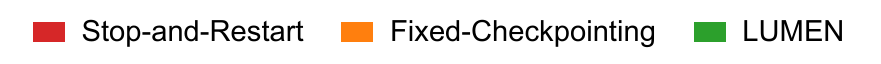}}
\vspace{-6pt}\\
\includegraphics[width=0.48\linewidth]{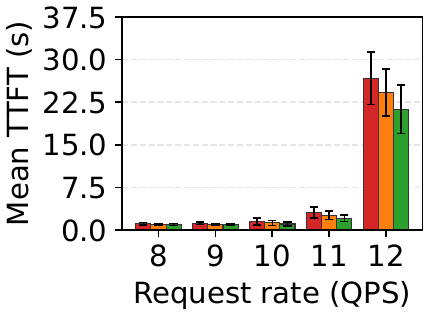} &
\includegraphics[width=0.48\linewidth]{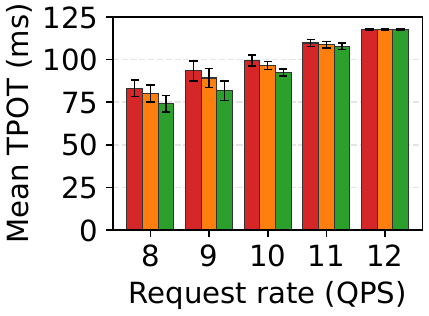} \\
{\small \makecell[c]{(a) Mean TTFT}} &
{\small \makecell[c]{(b) Mean TPOT}}
\end{tabular}
\vspace{-9pt}
\caption{Experiment~A.3 (Prototype): Impact of request rate.}
\label{fig:proto_qps}
\vspace{-6pt}
\end{figure}

\para{Experiment~A.3 (Impact of request rate).} Figure~\ref{fig:proto_qps} shows mean TTFT and mean TPOT across different request rates on the eight-worker deployment. Stop-and-Restart's mean TTFT increases from 1.5\,s at the default 10\,QPS to 26.7\,s at 12 QPS; at 8\,QPS and 9\,QPS, a single failure has limited impact, and the schemes differ by less than 0.2\,s. \sysname reduces mean TTFT by 29.6\% over Stop-and-Restart at the default 10\,QPS, with the reduction peaking at 33.3\% at 11\,QPS before saturation. Mean TPOT follows the same trend: \sysname remains the lowest through the speculation path as queues grow, reducing mean TPOT by 10.9\% at 8\,QPS and by 7.1\% at the default, while Stop-and-Restart's TPOT increases sharply.

\begin{figure}[!t]
\centering
\begin{tabular}{@{\ }c@{\ }c@{\ }}
\multicolumn{2}{c}{\includegraphics[width=0.95\linewidth]{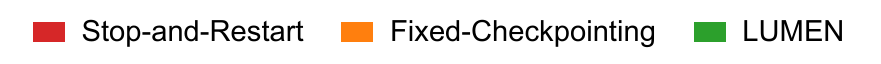}}
\vspace{-6pt}\\
\includegraphics[width=0.48\linewidth]{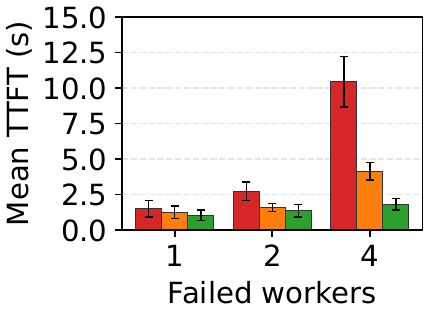} &
\includegraphics[width=0.48\linewidth]{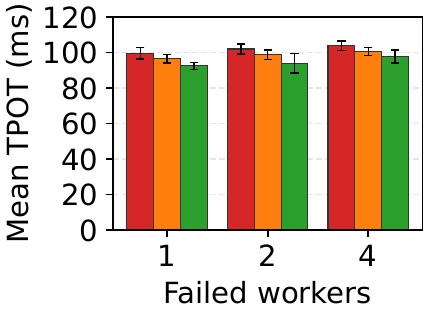} \\
{\small \makecell[c]{(a) Mean TTFT}} &
{\small \makecell[c]{(b) Mean TPOT}}
\end{tabular}
\vspace{-9pt}
\caption{Experiment~A.4 (Prototype): Impact of number of failures.}
\label{fig:proto_fcount}
\vspace{-6pt}
\end{figure}

\para{Experiment~A.4 (Impact of number of failures).} Figure~\ref{fig:proto_fcount} evaluates the impact of concurrent failures by failing one, two, and four out of eight workers. \sysname's gain grows with failure count, reducing mean TTFT over Stop-and-Restart by 29.6\%, 50.8\%, and 82.7\% for one, two, and four failed workers, respectively. Stop-and-Restart's mean TTFT increases sharply (1.5\,s at one failure to 10.4\,s at four) as its TPOT approaches the per-batch decode limit; \sysname scales more gracefully as speculation and KV reuse together eliminate the prefill replay that dominates under high failure counts.

\subsection{Simulator Evaluation}
\label{subsec:eval_simulator}

\begin{table}[!t]
\centering
\small
\caption{Experiment~B.1 (Simulator): End-to-end recovery.}
\label{tab:sim_e2e}
\vspace{-9pt}
\begin{tabular}{@{}lccc@{}}
\toprule
Scheme & Mean TTFT (s) & Mean TPOT (ms) & Recovery (s) \\
\midrule
S\&R & $1.6 \pm 0.34$ & $167.1 \pm 14.4$ & 100.3 \\
F-Ckpt & $1.5 \pm 0.26$ & $156.8 \pm 10.6$ & 94.9 \\
+Scheduling & $1.5 \pm 0.25$ & $156.7 \pm 9.7$ & 91.1 \\
+Progressive & $1.6 \pm 0.29$ & $136.4 \pm 9.9$ & 81.5 \\
\sysname & $1.4 \pm 0.20$ & $129.3 \pm 6.0$ & 81.5 \\
\bottomrule
\end{tabular}
\vspace{-6pt}
\end{table}

\noindent{\bf Experiment~B.1 (End-to-end recovery).} Table~\ref{tab:sim_e2e} reports end-to-end recovery on the 10-worker Llama-3-70B deployment under a single failure. \sysname reduces mean TPOT within the failure-impact window by 22.6\% over Stop-and-Restart and 17.6\% over Fixed-Checkpointing. Mean TTFT improvements are smaller (12.2\% and 5.1\% over the respective baselines) because a single failure across 10 workers raises TTFT only slightly while the cluster is unsaturated; this regime is consistent with the prototype at low load, and Experiment~B.2 confirms larger TTFT gains as load increases. \sysname also reduces recovery time by 18.7\% over Stop-and-Restart and 14.1\% over Fixed-Checkpointing. As recovery time is measured over the overall failure-impact window, the single-failure case is governed mostly by the many uninterrupted requests on surviving workers. Thus, +Progressive and \sysname have almost identical recovery time: speculation restores decode capacity during target-model reload, while the extra KV-reuse path affects only the small interrupted-request set. KV reuse still appears in mean TPOT (5.2\% below +Progressive), and its recovery-time benefit becomes visible when more workers fail (Experiment~B.3).

The breakdown in Table~\ref{tab:sim_e2e} reveals complementary roles. +Scheduling reduces mean TTFT by 6.3\% by eliminating lost-prefix replay, while +Progressive does not reduce mean TTFT in the rounded table (1.6\,s, matching S\&R) because the recovering worker cannot yet accept prefill while the draft model loads. For mean TPOT, roles are reversed: +Progressive contributes an 18.4\% reduction by handling decode through speculation, while +Scheduling contributes only 6.2\% because the recovering worker remains idle during target-model loading. \sysname combines both features and achieves the lowest mean TTFT and mean TPOT.

\begin{figure}[!t]
\centering
\begin{tabular}{@{\ }c@{\ }c@{\ }}
\multicolumn{2}{c}{\includegraphics[width=0.95\linewidth]{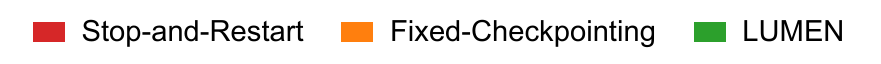}}
\vspace{-6pt}\\
\includegraphics[width=0.48\linewidth]{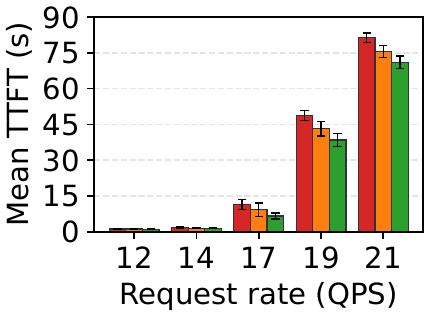} &
\includegraphics[width=0.48\linewidth]{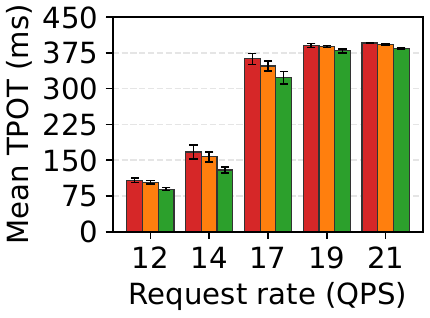} \\
{\small \makecell[c]{(a) Mean TTFT}} &
{\small \makecell[c]{(b) Mean TPOT}}
\end{tabular}
\vspace{-9pt}
\caption{Experiment~B.2 (Simulator): Impact of request rate.}
\label{fig:sim_qps}
\vspace{-6pt}
\end{figure}

\para{Experiment~B.2 (Impact of request rate).} Figure~\ref{fig:sim_qps} examines five request rates varied from 12 to 21\,QPS, spanning near-saturation (12--16\,QPS) and overload (17--21\,QPS). In steady state, \sysname's gain is led by speculation: mean TPOT is reduced over Stop-and-Restart by 17.2-22.6\% and over Fixed-Checkpointing by 14.2-17.6\% across 12--15\,QPS. Mean TTFT improves less at low load (5.3\% at 12\,QPS, 12.2\% at 14\,QPS over Stop-and-Restart), because a single failure barely perturbs an uncongested cluster. Under overload, the pattern reverses: growing queues amplify the TTFT gap (42.7\% below Stop-and-Restart at 17\,QPS), while the TPOT gap narrows as all schemes saturate.

\begin{figure}[!t]
\centering
\begin{tabular}{@{}c@{}c@{}}
\multicolumn{2}{c}{\includegraphics[width=0.96\linewidth]{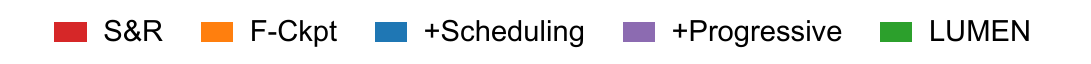}}
\vspace{-6pt}\\
\includegraphics[width=0.475\linewidth]{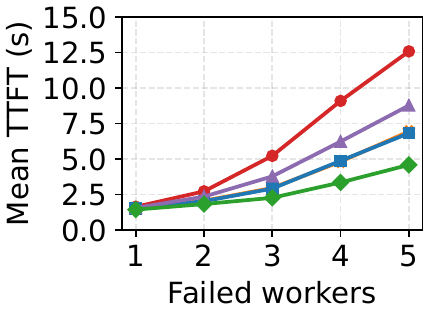} &
\includegraphics[width=0.475\linewidth]{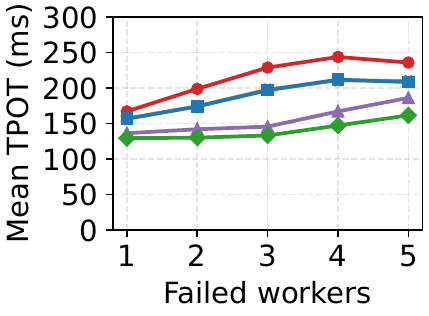}
\vspace{-3pt}\\
{\small \makecell[c]{(a) Mean TTFT}} &
{\small \makecell[c]{(b) Mean TPOT}}
\vspace{3pt}\\
\multicolumn{2}{c}{\includegraphics[width=0.95\linewidth]{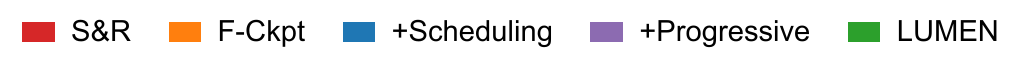}}
\vspace{-6pt}\\
\multicolumn{2}{c}{\includegraphics[width=0.95\linewidth]{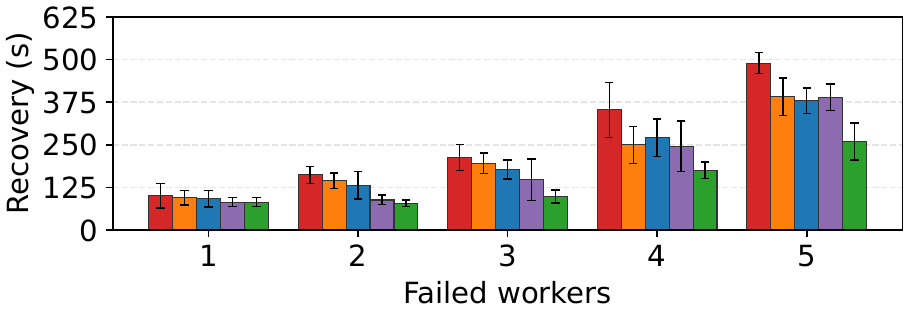}}
\vspace{-3pt}\\
\multicolumn{2}{c}{\small \makecell[c]{(c) Recovery time}}
\end{tabular}
\vspace{-9pt}
\caption{Experiment~B.3 (Simulator): Impact of number of failed workers on TTFT, TPOT, and recovery time.}
\label{fig:sim_fcount}
\vspace{-6pt}
\end{figure}

\para{Experiment~B.3 (Impact of number of failures).} Figure~\ref{fig:sim_fcount} varies one to five simultaneous failures across 10 workers. \sysname's advantage scales with failure count: at five failures, mean TTFT is 63.6\% below Stop-and-Restart and 33.9\% below Fixed-Checkpointing, mean TPOT is 31.5\% and 23.0\% lower, and recovery time is 46.9\% and 33.3\% shorter, respectively. With a single failure, the schemes remain close as interrupted requests account for a small fraction of all requests in the failure-impact window; recovery time is governed mainly by uninterrupted requests queuing behind redirected traffic on surviving workers. As the failure count grows, a larger share of requests lose their serving worker, increasing the impact on cluster-wide recovery. +Scheduling stays close to Fixed-Checkpointing because the default request rate leaves short queues, preventing the fixed checkpoint holder from becoming a bottleneck; the benefit of load-aware placement is more pronounced under higher load, as confirmed in Experiment~B.2.

\begin{figure}[!t]
\centering
\begin{tabular}{@{}c@{}c@{}}
\multicolumn{2}{c}{\includegraphics[width=0.96\linewidth]{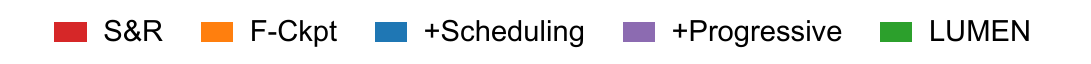}}
\vspace{-6pt}\\
\includegraphics[width=0.475\linewidth]{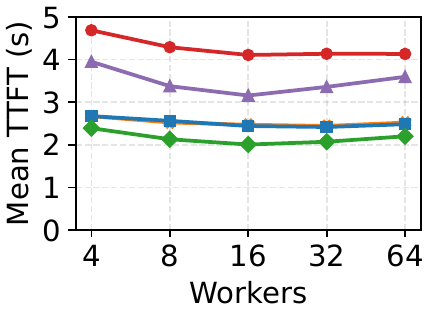} &
\includegraphics[width=0.475\linewidth]{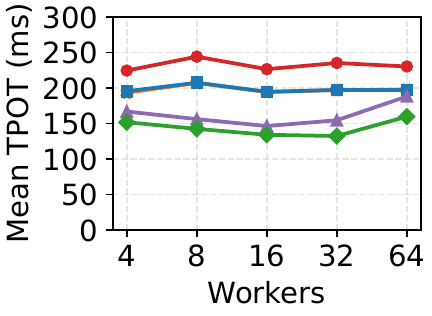}
\vspace{-3pt}\\
{\small \makecell[c]{(a) Mean TTFT}} &
{\small \makecell[c]{(b) Mean TPOT}}
\vspace{3pt}\\
\multicolumn{2}{c}{\includegraphics[width=0.95\linewidth]{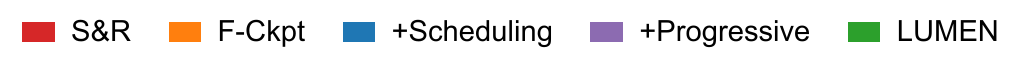}}
\vspace{-6pt}\\
\multicolumn{2}{c}{\includegraphics[width=0.95\linewidth]{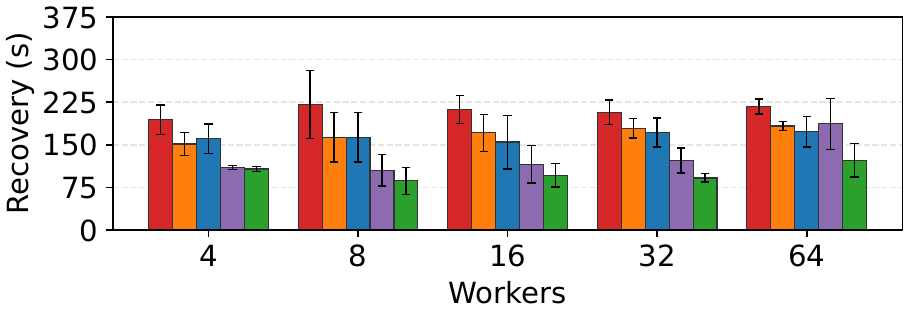}}
\vspace{-3pt}\\
\multicolumn{2}{c}{\small \makecell[c]{(c) Recovery time}}
\end{tabular}
\vspace{-9pt}
\caption{Experiment~B.4 (Simulator): Scalability with proportional failures on TTFT, TPOT, and recovery time.}
\label{fig:sim_wscale}
\vspace{-6pt}
\end{figure}

\para{Experiment~B.4 (Scalability with proportional failures).} Figure~\ref{fig:sim_wscale} scales the cluster from 4 to 64 workers with a fixed per-worker load of 1.4\,QPS and a fixed 25\% failure rate (i.e., 1--16 simultaneous failures). \sysname's gains are stable across all scales. Compared to Stop-and-Restart, mean TTFT stays 46.8--51.2\% lower at 4--64 workers and mean TPOT is 30.7--43.8\% lower over the same range; recovery time is shorter by 43.6--61.0\%. Both KV reuse and progressive recovery contribute: \sysname outperforms Fixed-Checkpointing by 12.8--18.8\% in mean TTFT and 19.2--32.9\% in mean TPOT at 8--64 workers, with the difference attributable to the speculation path that Fixed-Checkpointing lacks.  Note that under the fixed per-worker load, +Scheduling stays close to Fixed-Checkpointing because per-worker queueing pressure does not grow with cluster size, so the fixed checkpoint holder rarely becomes a bottleneck. \sysname's KV-reuse and progressive-recovery paths are therefore complementary: \sysname reduces recovery time by 29.0--47.0\% over +Scheduling (which lacks progressive recovery) and by 2.2--34.4\% over +Progressive (which lacks KV reuse). Furthermore, under 16 failures out of 64 workers, 23.8\% of interrupted requests fall back to full recomputation due to checkpoint-holder co-failure (\S\ref{subsec:discussion}).

\begin{figure}[!t]
\centering
\begin{tabular}{@{\ }c@{\ }c@{\ }}
\multicolumn{2}{c}{\includegraphics[width=0.95\linewidth]{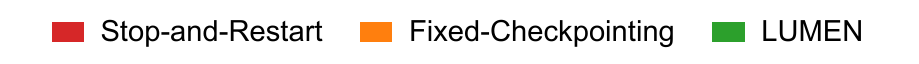}}
\vspace{-6pt}\\
\includegraphics[width=0.48\linewidth]{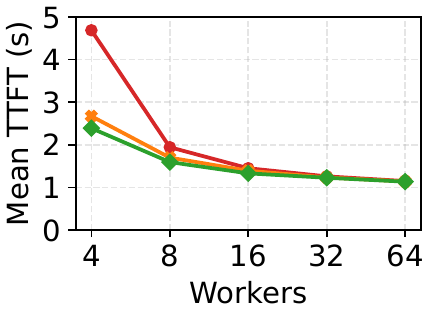} &
\includegraphics[width=0.48\linewidth]{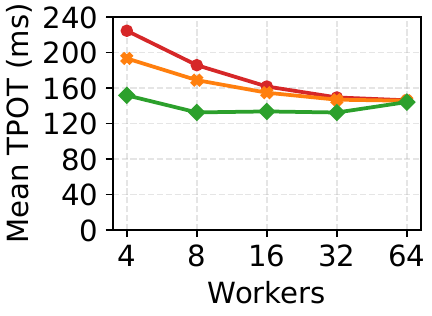}
\vspace{-3pt}\\
{\small \makecell[c]{(a) Mean TTFT}} &
{\small \makecell[c]{(b) Mean TPOT}}
\end{tabular}
\vspace{-9pt}
\caption{Experiment~B.5 (Simulator): Scalability with a single failure.}
\label{fig:sim_wscale_single}
\vspace{-6pt}
\end{figure}

\para{Experiment~B.5 (Scalability with a single failure).} Figure~\ref{fig:sim_wscale_single} varies the cluster size while keeping a single failure, so the failure share decreases as the cluster size increases. \sysname has the lowest latency at every scale, but its benefit shrinks as the single failure becomes a smaller fraction of the cluster. Stop-and-Restart's mean TTFT within the window falls from 5.2\,s at four workers to 1.2\,s at 64, whereas under the fixed 25\% failure fraction it remains within 4.1--5.2\,s (Figure~\ref{fig:sim_wscale}). As a result, \sysname's TTFT reduction over Stop-and-Restart falls from 50.6\% at four workers to 1.3\% at 64, and the TPOT reduction declines from 37.2\% to 3.7\% over the same range. Together, Experiments~B.4 and B.5 span the full range of operating conditions: \sysname removes much of the recovery cost when the failed fraction is high, and does not increase latency when a single failure is a negligible share of a large cluster. Table~\ref{tab:sim_wscale_reqtype} further confirms that \sysname reduces TPOT for interrupted requests by 53--67\% over Stop-and-Restart across all scales through speculation.

\begin{table}[!t]
\centering
\small
\caption{Experiment~B.5: Per-request-type breakdown across worker counts.}
\label{tab:sim_wscale_reqtype}
\vspace{-9pt}
\resizebox{\columnwidth}{!}{\begin{tabular}{@{}cl ccc ccc@{}}
\toprule
\multirow{2}{*}{Workers} & \multirow{2}{*}{Request type} & \multicolumn{3}{c}{Mean TTFT (s)} & \multicolumn{3}{c}{Mean TPOT (ms)} \\
\cmidrule(lr){3-5} \cmidrule(lr){6-8}
 & & S\&R & F-Ckpt & \sysname & S\&R & F-Ckpt & \sysname \\
\midrule
\multirow{2}{*}{4} & Interrupted & 27.51 & 23.61 & 23.44 & 357.7 & 176.9 & 116.5 \\
 & Uninterrupted & 4.10 & 2.14 & 1.85 & 231.6 & 200.9 & 154.6 \\
\midrule
\multirow{2}{*}{8} & Interrupted & 24.26 & 22.26 & 22.08 & 236.1 & 115.7 & 79.6 \\
 & Uninterrupted & 1.55 & 1.34 & 1.24 & 177.2 & 163.8 & 133.8 \\
\midrule
\multirow{2}{*}{16} & Interrupted & 22.08 & 21.20 & 20.61 & 177.7 & 129.9 & 76.4 \\
 & Uninterrupted & 1.25 & 1.20 & 1.14 & 157.9 & 151.8 & 135.6 \\
\midrule
\multirow{2}{*}{32} & Interrupted & 26.28 & 25.55 & 25.36 & 150.0 & 87.8 & 60.0 \\
 & Uninterrupted & 1.11 & 1.09 & 1.09 & 147.1 & 143.8 & 133.8 \\
\midrule
\multirow{2}{*}{64} & Interrupted & 23.43 & 22.60 & 22.47 & 147.6 & 82.0 & 69.6 \\
 & Uninterrupted & 1.05 & 1.05 & 1.04 & 141.7 & 140.5 & 136.9 \\
\bottomrule
\end{tabular}
}
\vspace{-6pt}
\end{table}

\begin{table}[!t]
\centering
\small
\caption{Experiment~B.6 (Simulator): \sysname's TTFT and TPOT across speculative depth $K$, each paired with its acceptance rate $\alpha$.}
\label{tab:sim_kdepth}
\vspace{-9pt}
\resizebox{0.80\columnwidth}{!}{%
\begin{tabular}{lccc}
\toprule
 & $K=2$ & $K=4$ & $K=8$ \\
\midrule
Acceptance rate $\alpha$ & 0.72 & 0.60 & 0.50 \\
Mean TTFT (s) & $1.4 \pm 0.20$ & $1.4 \pm 0.20$ & $1.4 \pm 0.20$ \\
Mean TPOT (ms) & $130.1 \pm 6.2$ & $129.3 \pm 6.0$ & $128.9 \pm 5.9$ \\
\bottomrule
\end{tabular}}
\vspace{-6pt}
\end{table}

\para{Experiment~B.6 (Speculative-depth sensitivity).} Table~\ref{tab:sim_kdepth} varies $K$ over (2, 4, 8) draft tokens per step using the measured acceptance rates (0.72, 0.60, 0.50) for the Llama-3-70B pair. \sysname remains insensitive to $K$: its mean TTFT stays at 1.4\,s and mean TPOT changes by only 0.9\% across $K$, both within confidence intervals. This insensitivity is inherent to speculative decoding: a deeper draft proposes more tokens per step, but the correspondingly lower acceptance rate cancels the gain, keeping the expected number of accepted tokens near constant (around 1.2, 1.3, and 1.0 under the respective acceptance rates) while each verification step becomes more expensive. The default $K = 4$ is robust.

\begin{table}[!t]
\centering
\small
\caption{Experiment~B.7 (Simulator): \sysname's TTFT and TPOT across the checkpoint-placement weight $\lambda$.}
\label{tab:lambda_sensitivity}
\vspace{-9pt}
\resizebox{\columnwidth}{!}{%
\begin{tabular}{lccccc}
\toprule
$\lambda$ & 0.25 & 0.5 & 1 & 2 & 4 \\
\midrule
TTFT (s) & $1.4 \pm 0.20$ & $1.4 \pm 0.20$ & $1.4 \pm 0.20$ & $1.4 \pm 0.20$ & $1.4 \pm 0.21$ \\
TPOT (ms) & $129.0 \pm 5.9$ & $129.0 \pm 6.0$ & $129.0 \pm 5.7$ & $129.1 \pm 5.7$ & $129.2 \pm 5.9$ \\
\bottomrule
\end{tabular}}
\vspace{-6pt}
\end{table}

\para{Experiment~B.7 (Checkpoint-placement weight sensitivity).} Table~\ref{tab:lambda_sensitivity} varies $\lambda$ (Equation~\eqref{eq:placement}) from 0.25 to 4. \sysname's mean TTFT stays at 1.4\,s and mean TPOT remains within 129.0--129.2 ms across $\lambda$ (under 0.5\% variation, within confidence intervals). Under a single-worker failure, $\lambda$ only affects where the KV checkpoints of the few interrupted requests are placed; mean TTFT and TPOT within the failure-impact window are governed primarily by the surviving workers absorbing the failed worker's traffic. The default $\lambda = 1$ is robust for single-failure conditions.

\subsection{Highlights of Evaluation Results}

\sysname's performance gains over Stop-and-Restart and Fixed-Checkpointing are most pronounced under three conditions: (i) high failure severity, where mean TTFT reductions over Stop-and-Restart grow from 29.6\% at one failed worker to 82.7\% at four on the prototype (Experiment~A.4) and from 12.2\% to 63.6\% on the simulator (Experiment~B.3); (ii) high cluster load, where near-saturation queues amplify the TTFT advantage to 42.7\% over Stop-and-Restart (Experiment~B.2); and (iii) proportionally large failure fractions, where gains remain stable at 46.8--51.2\% below Stop-and-Restart in mean TTFT across 4 to 64 workers at a fixed 25\% failure rate (Experiment~B.4).
In all conditions, \sysname adds no measurable steady-state overhead.

\section{Related Work}
\label{sec:relatedwork}

\noindent{\bf Fault tolerance in LLM training.}
Gemini \cite{wang23} shortens training recovery time by storing in-memory checkpoints of model weights and optimizer states on other workers' hosts. Bamboo \cite{thorpe23}, Oobleck \cite{jang23}, and Recycle \cite{gandhi24} achieve resilient training through redundant pipeline execution and template-based reconfiguration. ECCheck \cite{qi25} lowers the memory cost of in-memory checkpoints through erasure coding. ByteCheckpoint \cite{wan25} unifies large foundation-model training checkpointing across parallelism strategies. However, the above training studies target long-running training jobs; they do not address per-request state under interactive latency requirements as \sysname does.

\para{Fault tolerance in LLM serving.}
Recent work makes LLM serving resilient to GPU failures by reusing previously computed KV cache. D\'ej\`aVu \cite{strati24} streams KV cache to another worker's CPU memory through a fixed ring topology and always assigns each worker's checkpoint to a fixed neighbor. FailSafe \cite{xu26} sustains intra-node serving under irregular GPU availability, while GhostServe \cite{jayakody26} protects KV cache with erasure-coded parity shards in host memory.
KevlarFlow \cite{qian26} addresses multi-worker recovery through GPU-memory KV replication and pipeline reconfiguration, but does not route recovery in a load-aware manner.
\sysname instead distributes recovery work using recent worker-load observations and overlaps full-model loading with draft-generation assistance to keep the recovering worker productive.

\para{Efficient LLM scheduling.}
Orca \cite{yu22} proposes iteration-level scheduling for transformer serving. vLLM \cite{kwon23} introduces paged KV management, which \sysname builds on for request-state movement and reuse. Sarathi-Serve \cite{agrawal24sarathi} improves inference efficiency with chunked prefills and decode piggybacking. SGLang \cite{zheng24} provides an efficient execution substrate for structured language model programs, and \sysname builds directly on this runtime. Llumnix \cite{sun24} supports live request migration for dynamic scheduling.
FastServe \cite{wu26} introduces preemptive scheduling with proactive job swapping to minimize latency. However, all the above studies assume worker availability in the steady state and do not enable a recovering worker to serve before its model is fully reloaded. \sysname's load-aware recovery is complementary to these schedulers without modifying their steady-state scheduling logic.

\section{Conclusion}
\label{sec:conclusion}

We study the failure recovery problem for distributed LLM serving, where worker failures lead to the loss of both in-flight KV cache and serving capacity, thereby degrading both TTFT and TPOT. \sysname is a fault-tolerant LLM serving system that makes {\em load-aware} recovery decisions. It incorporates three novel techniques: load-aware KV checkpointing, locality-aware recovery scheduling, and speculation-assisted progressive recovery. Our prototype and simulation results show that \sysname reduces TTFT, TPOT, and recovery time compared to the recovery baselines Stop-and-Restart and Fixed-Checkpointing across various settings.

\end{sloppypar}

\bibliographystyle{plain}
\bibliography{ref}

\end{document}